\documentclass{article}
\usepackage{cite}
\usepackage[multiple]{footmisc}
\begin{document}
\begin{titlepage}
\begin{center}
\large
\textbf{Mind and Matter:  A Critique of Cartesian Thinking}
\end{center}
\vspace{0.5 cm}
\begin{center} Marcus Appleby
\\
\vspace{0.1 in}
\emph{Perimeter Institute for Theoretical Physics,  Waterloo, Ontario N2L 2Y5, Canada} \\  \vspace{0.05 in} \emph{and}  \vspace{0.07 in}  \\
\emph{Stellenbosch Institute for Advanced Study, Stellenbosch, Matieland 7602, South Africa}
 \end{center}
\vspace{1 cm}
\begin{center} \textbf{Abstract}

\vspace{0.5 cm}

\parbox{12 cm }{

}
\vspace{0.35 cm}
\parbox{12 cm }{It is argued that the problem of interpreting quantum mechanics, and the philosophical problem of consciousness, both have their roots in the same set of misguided Cartesian assumptions.  The confusions underlying those assumptions are analyzed in detail.   It is sometimes suggested that quantum mechanics might explain consciousness.  That is not the suggestion here.  Rather it is suggested that an adequate non-Cartesian philosophy would transform our understanding of both quantum mechanics and consciousness.  Consequently, it would change our ideas as to just what it is that we are trying to explain.}
\end{center}
\end{titlepage}

\section{Introduction}
Pauli, in a letter to van Franz (quoted Gieser~\cite{Gieser}, pp.243--4), wrote
\begin{quote}
Evidently the progress of science must take such a course that the concept `consciousness' will be replaced by a more general or better one.
\end{quote}
If one knew that these words were written by a leading $20^{\mathrm{th}}$ scientist, but did not know that the scientist in question was Pauli, one might think that what is being advocated here is eliminative materialism, or some such similar position (refs.~\cite{StichA,StichB,DennettA,DennettB,ChurchlandA,ChurchlandB}, and references cited therein).  Since, however, it is Pauli who is saying this we know he must be thinking along very different lines.  Eliminative materialists propose to deal with the mind-body problem by eliminating the mental pole of the duality leaving only the material one. Pauli would reject that proposal  because he was looking, not for a materialistic explanation of mental phenomena, but rather for a ``psychophysical monism"  in which mind and matter are seen as ``two aspects of one and the same abstract fact'', itself neither physical nor psychological (Meier \emph{et al}~\cite{JPLetters}, pp.87, 159).   It is easy to see why a materialist might want to take an eliminativist attitude to consciousness.  The question addressed in this paper is why someone like Pauli, who is not a materialist, would take such an attitude.

What follows is not an exercise in Pauli exegesis.  I am not here particularly concerned with Pauli's  reasons for taking that view of consciousness.  Rather, I am going to give my  own reasons for thinking that he might have been basically right.

Before proceeding further, I ought to qualify.  The meaning of a word like ``cat'', which can be defined ostensively, is securely anchored.  However, the word ``consciousness'' cannot be defined ostensively, not even by the person whose consciousness it is  (it is surely not possible to point one's finger at one's own consciousness).  Consequently, if one is not careful, there is a danger that its meaning will float, so that it comes to be used in different ways by different people, or even by the same person at different times.  I believe this actually happens.  The criticisms of this paper are only directed at \emph{one} of its possible senses.

As an example of a sense of the word which I feel  is unlikely to be rendered obsolete by future scientific advance, consider the Glasgow Coma Scale~\cite{TeasdaleA,TeasdaleB} which is widely used  to quantify the level of consciousness in cases of brain damage.  It is possible, even likely, that the Glasgow Coma Scale will, in time, come to be replaced by some improved method for quantifying degree of consciousness.  It is also likely that scientific advances will lead to a deeper and richer understanding of the phenomenon itself.  However, I doubt that this would amount to the kind of development Pauli had in mind when he wrote of the concept of consciousness being ``replaced by a more general or better one''. 

For want of a better term I will refer to the sense in which the word ``consciousness'' is used in medicine as its ``everyday sense''.  It is true that the medical literature on the subject can be quite technical.  However, although medical science has refined the description of states of consciousness, it has done so in a way which remains close to the root meaning.  A doctor will understand the statement ``the patient is fully conscious'' in almost, if not exactly the same sense that the patient's relatives understand it. I take the everyday sense of the word also to include its use in sentences like ``She was conscious of the clock ticking,'' to describe the state of being aware of something.

The critical comments in this essay are directed, not at consciousness in the everyday sense, but rather at the concept as it is used in, for example, philosophical discussions of the so-called problem of consciousness.  I will refer to this second sense of the word as the Cartesian sense.  It is true that nowadays there are not many full-blooded Cartesian dualists left.  Nevertheless, a more or less attenuated version of the Cartesian soul continues to be prominent in modern philosophical thinking, and it is this which gives rise to the ``problem of consciousness''.   It is clear from context\footnote{In particular, it is clear that Pauli had in mind the so-called privacy of Cartesian consciousness---the property of being undetectable by the outside observer.}  that it was Cartesian consciousness that Pauli had in mind when he made the statement quoted at the beginning of this essay.

To see  that the everyday and Cartesian senses are different consider the discussion in Chalmers~\cite{Chalmers}.  Chalmers begins by saying that consciousness is ``intangible'' and consequently  hard to define (p.3), which I think is already an indication that what is in question is something different from consciousness in the everyday sense (consider the likely response of a hospital doctor  to the proposition that the state of  being non-comatose is intangible, and hard to define).     He then goes on to propose the characterization ``the subjective quality of experience'' (p.4).  Now the meaning of this   will be clear enough to someone who has received a certain kind of education.  More specifically, it will be clear to someone who has absorbed the basic ideas of the Cartesian philosophy.  But I believe  it would be unintelligible to anyone  who has not had the benefit of such an education (probably the majority of English speakers).  

What Chalmers thinks of as the subjective quality of greenness, philosophically unsophisticated people think of simply as greenness, and it would take a lot of work to persuade them that they are missing something important.  Something that is not taken for granted by the vast majority of speakers cannot be considered to belong to the everyday sense of a word.  Of course, one might think that the Cartesian concept of consciousness can be seen to be logically contained in  everyday assumptions, if one takes the trouble to think the matter through carefully.  However,  it is precisely the point of this paper that it is \emph{not} so contained. 

Chalmers, like others, thinks that consciousness is hard to define.  Why should that be?  I believe that Searle~\cite[p.121]{Searle} puts his finger on at least part of the difficulty when he says
\begin{quote}
The reason we find it difficult to distinguish between my description of the objects on the table and and my description of my experience of the objects is that the features of the objects are precisely the conditions of satisfaction of my conscious experiences of them.   So the vocabulary I use to describe the table---``There's a lamp on the rich and a vase on the left and a small statue in the middle''---is precisely that which I use to describe my conscious visual experiences of the table.  (p.131)
\end{quote}
Which provokes the obvious question:  if two things have the same description, how does one tell them apart?  \emph{Can} one tell them apart?  Could it just be that what 
 Searle seeks to convey by the phrase ``the contents of my consciousness when I look at my table'' is identical to what a less sophisticated person would convey more succinctly, simply by saying ``my table''? It seems, however, that that cannot be precisely right, for Searle  argues that consciousness is always perspectival.  Consequently, he thinks that his visual consciousness of his table only comprises the parts he can directly see.  Nevertheless, it is  hard to resist the impression that what Searle means by the phrase ``the contents of my consciousness'' is, if not identical, at any rate close to what an unsophisticated person means by the phrase ``the things around me'':  that the contents of Searle's consciousness, as Searle conceives them to be, can be pictured as something like a film set, convincing when seen from the front, unpainted wood when seen from the back.  

This way of thinking is historically important, because it led to idealism.  In an amusing critique of idealist philosophy Stove~\cite[p.116]{Stove} asks what is the ``product-differentiation'':   \emph{i.e.} ``what are they \emph{selling}, these people who call themselves objective idealists, that a commonsense \emph{materialist} could not consistently buy?''.  His answer is that there is in fact nothing that a  materialist could not consistently buy.   In support of this conclusion he cites Bosanquet (one of the more prominent $19^{\mathrm{th}}$ century idealists), who said in so many words that ``extremes meet'', and ``a consistent materialist and thorough idealist hold positions which are distinguishable only in name'' (\emph{ibid}, p.115).  

These days idealism has gone out of fashion.  However, believers in Cartesian consciousness are still faced with what is essentially the same problem, of differentiating the contents of  consciousness (as they conceive them to be) from what commonsense would call the objects around us.  It is a difficult problem, and I think that is one of the reasons it is often said that ``consciousness'' is hard to define. 

\section{The Cartesian Split}

Descartes introduced a fundamental split between Cartesian consciousness and  Cartesian matter.  I am here using the term ``Cartesian matter'' rather loosely, to refer, not only to the concept of matter originally proposed by Descartes himself, but also to its many descendants.   I described the concept of consciousness as it features in, for example, the book by Chalmers~\cite{Chalmers} as an attenuated variant of the Cartesian soul.  In the same way I would, for example, describe  the universal wave function proposed by Everett\cite{Everett}  as  a (not so attenuated) variant of Cartesian matter.  It goes without saying that Chalmers' concept of consciousness differs greatly from Descartes' concept of the soul.  However, it shares with the latter the crucial feature of being a receptacle for  all the supposedly subjective phenomena which, on a Cartesian view, are excluded from the physical universe.  Similarly,  Everett's concept of the universal state vector, though obviously very different from  Descartes' concept of matter, still shares the crucial feature, that it is supposed to be completely describable in purely objective, mathematical terms, without any contamination by the observing subject. 

The point to notice is that these two concepts,  Cartesian consciousness and Cartesian matter, are different aspects of a single conceptual scheme.  They are like the two poles of a bar magnet, impossible to isolate.  Idealists attempt to cut the bar in two, keeping only the subjective side of the polarity.  But, as we saw, when they try to carry that idea through consistently it turns out that the concept of matter has come back in, through the backdoor, so to speak.  Materialists attempt to perform the same bisection, keeping only the objective side of the polarity.  However, they then face the problem that, no matter how vigorously they attempt to cast doubt on the notion of \emph{qualia}  (see, for instance, \cite[section 17]{Lycan}), the fact remains that, to a normally sighted person, green things  undeniably do look qualitatively different from red ones.  Consequently, if one looks at a green object, while trying  to keep in mind that the quality of perceived greenness is not really a feature of the object itself, it is  difficult to avoid the thought that the quality of greenness is a feature that is somehow added by one's own perceptual apparatus. From there it is but a small step to the Cartesian concept of consciousness.  

I believe we need to break away from this whole misguided way of thinking:  not simply to deny Cartesian consciousness, nor simply to deny Cartesian matter, but  to deny both. There  are many empirical reasons for taking such a course.  Modern neuroscience gives us  reasons for being   suspicious of Cartesian assumptions about consciousness (see~\cite{DennettA,DennettB,BlackmoreC,BlackmoreD}, and references cited therein); while quantum mechanics gives us equally good reasons for being suspicious of Cartesian assumptions about matter (see any textbook).  

The aim of  physics, as Descartes conceived it, is to arrive at the one true picture of things,  totally objective, and complete in every detail.  Before the year 1900 it might have looked as though we were getting steadily closer to that goal\footnote{Although there were  $19^{\mathrm{th}}$ century physicists, such as Mach\cite{Mach}, who did not agree with Descartes about the goal of physics.}. However,  quantum mechanics strongly suggests that the goal is unachievable.  In quantum mechanics what you see depends on how you look.  Make one kind of measurement on the electromagnetic field and one will obtain results consistent with it being a smoothly varying wave; make another, different kind of measurement and one will obtain results consistent with it being a collection of discrete particles.   Similarly, if one observes an atom using a scanning tunnelling electron microscope, one will see an apparently solid object; If, on the other hand, one observes it with a $\gamma$-ray microscope one will see a collection of point-like particles separated by empty space.  

So which of these pictures is the \emph{true} one?  Quantum mechanics declines to say, just as it declines to say what is going on in a physical system when no one is looking.   In place of the God-like conspectus of the entire universe, with nothing left out, which Descartes imagined and which continued to inspire physicists for $250$ years after him, quantum mechanics merely gives us methods for anticipating what will be observed in this or that particular experimental context.  Moreover, the fact, that the  outcome depends on the observer's decision as to which  measurement to make, casts doubt on the assumption, that physics passively records events that would have happened anyway, in the absence of  experimental intervention.  This represents a subtle, but important departure from the Cartesian ideal of total objectivity.  

Since the 1920s there have been numerous attempts to reconcile quantum mechanics with Cartesian assumptions, as to what the world \emph{ought} to be like (for an overview see, for example, Schlosshauer~\cite{SchlosshauerB}).  These attempts have been successful to the extent that it seems there is nothing to logically exclude the possibility that, underlying the observations, there is some universal mathematical mechanism.  The difficulty is finding a picture of this kind which is empirically substantiated.    

When Einstein embarked on the project, of finding an alternative to the Copenhagen Interpretation, he doubtless hoped to find a single theory which, like the general theory of relativity, would be uniquely specified by the interplay of various empirical and  aesthetic considerations.  Doubtless he also hoped  for new empirical predictions.    Of course, conclusive demonstrations are not to be had in science.  So no one can say for sure that Einstein's hopes will not be fulfilled at some time in the future.  But it does seem to me that the effect of eighty years of theoretical work has been to make those hopes look increasingly forlorn.

My own feeling is that an adequate understanding of quantum mechanics ultimately depends, not on sophisticated technical developments, but on some simple conceptual shift---something a little like the perceptual shift which occurs when one looks at a diagram like the Necker cube, or the duck-rabbit picture (Wittgenstein~\cite{WittgensteinB} p.194e,  Kihlstrom~\cite{Kihlstrom}).  I doubt that quantum mechanics is   intrinsically weird.  It only seems weird because we insist on looking at it through Cartesian spectacles.  The problem is that Cartesian assumptions have become so deeply ingrained in our thinking that it is hard to find the right non-Cartesian spectacles.  

Turning to the other pole of the Cartesian duality, philosophers are familiar with the privacy of Cartesian consciousness:   the fact that  the  consciousness of another person is, from the Cartesian point of view, just as inescapably hidden as  the wave function is  in the Bohm interpretation of quantum mechanics (Bell~\cite{Bell}, p.202).  What is less widely appreciated  is that there is a problem with ascertaining the contents of  one's \emph{own} consciousness. 

 A particularly striking illustration of this point comes from the study of eye movements in reading~\cite{Rayner,RaynerB}.  In order to explain it I first need to say something about the physiology of  human vision. The region of the retina where the receptors are packed most tightly, and where visual acuity is consequently highest, is called the fovea.  The part of the visual field which falls on the fovea subtends an angle of $\sim 1^{\circ}$ at the centre of the lens.  Visual acuity falls off  rapidly as one moves away from this region, which means that in a single fixation of the eyes one is  able to discriminate fine  detail in only a very small portion of the visual field (a portion about the size of a thumbnail held at arm's length).  

The reason the visual system is nonetheless able to acquire accurate information about the whole environment is that the eyes are continually performing jumps, or saccades.  When reading the duration of a single saccade is typically  $\sim 30\ \textrm{ms}$, while the duration of the fixation between saccades is typically $\sim 200\ \textrm{ms}$ (in other activities the saccades are often bigger, and take correspondingly longer).  During a saccade very little information is transmitted to the cortical processing areas (this phenomenon is called saccadic suppression, or saccadic masking).  It can consequently be said that most of our visual awareness is based on $~\sim 4$ snapshots per second, each of them covering only a  small fraction of the visual field.  

These facts  already seem very counter-intuitive from a Cartesian point of view:  it is surprising (on Cartesian assumptions) that at any moment one sees so little in fine detail, and suprising also that there are so few jumps per second (a movie which ran at 4 frames per second would  look jumpy).    However, it gets worse (from a Cartesian point of view).  The eye muscles give a brief twitch to initiate a saccade, and thereafter the eyeballs move ballistically, subject only to frictional forces.  Consequently, a computer attached to an eye-tracking device can calculate where the next fixation is going to be before the eyes actually land there.  

This makes possible the following experiment.  One takes a page of printed text and projects it onto a screen, replacing all the letters by x's.  The experimental subject sits in front of the screen, and his/her eye-movements are monitored.  During a saccade the computer calculates where the eyes are going to alight, and puts  a handful of letters from the original page just at that point, leaving x's everywhere else.  In the next saccade the computer wipes those letters, replacing them by x's, and puts another group of letters at the next fixation point.  And so on.  To illustrate, in one experiment the original text was~\cite{Rayner}
\begin{quote}
\begin{center} 
By far the single most abundant substance in the biosphere 
\\
is the familiar but unusual inorganic compound called water. In
\\
nearly all its physical properties water is either unique or at
\\
the extreme end of the range of a property. It's extraordinary
\end{center}
\end{quote}
while what appeared on the screen during one particular fixation was
\begin{quote}
\begin{center}
\small
Xx xxx xxx xxxxxx xxxx xxxxxxxx xxxxxxxxx xx xxx xxxxxxxxx \\
xx xxx xxxxxxxx xxx xxxsual inorganic coxxxxxx xxxxxx xxxxx. Xx \\
xxxxxx xxx xxx xxxxxxxx xxxxxxxxxx xxxxx xx xxxxxx xxxxxx xx xx \\
xxx xxxxxxx xxx xx xxx xxxxx xx x xxxxxxxx. Xx'x xxxxxxxxxxxxx
\end{center}
\end{quote}
  This, and other, similar techniques have been used to acquire a wealth information about the visual system.  However, its relevance to the present discussion is simply this.  To an observer whose eye movements are not synchronized with the screen it is obvious (a)  that at any moment the screen contains almost nothing but x's and (b) that  what is on the screen is constantly changing.  However, to the experimental subject, whose eye movements \emph{are} synchronized, the screen looks like a perfectly normal page of text. To convey just how good the illusion is Grimes~\cite{Grimes} records that one of the first people to conduct an experiment of this kind served as the first experimental subject; after a while he sat back from the apparatus and announced that something must be wrong with the system because the text was not changing---though it was, in fact, working perfectly  (also see \cite[p.361]{DennettA}).

If one reflects on this fact, that it is demonstratively impossible to tell the difference between a normal page of printed text, and a page which at any given moment consists almost entirely  of x's, then one becomes genuinely uncertain, as to what precisely  \emph{are} the contents of one's own consciousness at any given moment.  Looking at the page in front of me I can see that it does not consist almost entirely of x's.  I am able to know this  because information is integrated across saccades.  Consequently, I am aware, not only  of the information  acquired on this present visual fixation, but also of information acquired on many previous fixations.  But \emph{how much} information is integrated across saccades? What precisely is its nature?  And precisely how much of that information is contained in my consciousness?  

The first two of these questions are empirical questions which can be, and actually are being investigated by the usual scientific methods.  However, the last is of a different character.  At least, it is of a different character if it is consciousness of the Cartesian sort which is in question.  On Cartesian principles, consciousness is private.    It follows that if I myself cannot tell what exactly are the contents of my own  consciousness, then no amount of neuroscientific experimentation can tell either.  That is the case for my consciousness of the printed text now in front of me.   Like the position of the particle in a two-slit experiment, my consciousness  now is indeterminate.

There are numerous other experiments and examples pointing to the same conclusion\footnote{See~\cite{BlackmoreC,BlackmoreD,DennettA,DennettB,Grimes,Simons}, and references cited therein}.  I will  confine myself to just two other examples.  Grimes~\cite{Grimes} used an eye-tracking device coupled to a computer to examine what happened when a picture (as opposed to a page of printed text) was changed in the middle of a saccade.  In one such experiment, in a picture of two men wearing differently coloured hats, the hats were switched mid-saccade.  100\% of the experimental subjects did not notice.  Even more dramatically, in another case a parrot, occupying roughly 25\% of the picture area, was switched from brilliant green to brilliant red mid-saccade.  In this case most of the subjects did notice.  But 18\% of them did not.  25\% of the picture area is a lot, and it raises the question:  what exactly is one conscious of, if one does not notice a change as striking as that? 

 A second illustration is the one given by Dennett~\cite[pp.354-5]{DennettA} (pp.354-5), of wallpaper in which the pattern consists of a large number of identical images of Marilyn Monroe.  If one looks at it it will only take one a second or two to realize that the images are all the same.  Since the eye performs only a few saccades per second it is impossible that one has discriminated more than a handful of the images in sufficient detail to be able to identify it.  Instead the visual system must essentially be making a guess, based on the small number of cases which it has accurately discriminated.  So the question arises again:  in a case like this what exactly are the contents of consciousness? 

In ordinary life, and in physics also before the $20^{\mathrm{th}}$ century, the assumption, that a physical object always has a determinate trajectory, works very well.  But when we push our investigations far enough we start to run into difficulties.  Similarly with the concept of consciousness:  when we start to ask the kind of detailed questions raised in the last few paragraphs we run into problems which are not entirely dissimilar to the problems which quantum mechanics reveals for Cartesian matter.

It is often thought that quantum indeterminacies are humanly un\-imaginable.  That is to get it exactly the wrong way round.  What is impossible to imagine is knowing the position of something to infinitely many decimal places.  On other hand, ordinary experience is full of indeterminacies.  If someone wants to know what it would be like to perceive an indeterminate position all they need do is  look at an object in a room, and  try to estimate its distance from the walls.  It is unlikely that they can achieve even 10\% accuracy.  Similarly, to know what it is like to perceive a number indeterminacy (such as the indeterminacy of photon number in a coherent state) all one need  do is   look at a collection of objects on a table.  If one is then asked how many objects there are it is unlikely one will be able to say, without first taking the time to count them up.  The fact that one cannot answer straight away (and probably could not answer at all if one did not still have the objects in view) suggests that at the time of asking one was conscious of the objects, but not of their number.

Dennett has written a book entitled \emph{Consciousness Explained}~\cite{DennettA}.  Since I agree with Dennett on a number of  points I ought to stress that I do not agree with him on this central one.  Specifically, I do not think that  he, or anyone else, is  close to ``explaining  consciousness''.  Like Pauli, I think that a satisfactory understanding of these questions will involve breaking out of the Cartesian mould entirely, and developing a different conceptual framework.

At this stage I should perhaps obviate another potential misunderstanding.  There have been a number of attempts to explain consciousness using quantum mechanics (see Atmanspacher~\cite{AtmanspacherC} for a review).  Since  these approaches all depend on adopting a non-Copenhagen interpretation of quantum mechanics, and since they take the Cartesian concept of consciousness for granted, it should be apparent, from what I said earlier, that I do not find any of them convincing.  If I keep mentioning consciousness and quantum mechanics in the same breath (so to speak) it is not because I think that one of them can be used to explain the other, but because I think that in both cases a clear understanding of the phenomena is obstructed by the same misguided Cartesian philosophy.  A second, subsidiary reason is that I cannot help being struck by parallels\footnote{For other discussions of this, and related points see refs.~\cite{JPLetters,Jung,AtmanspacherA,AtmanspacherB,Conte,Bruza}, and references cited therein.}.  What the parallels are worth, I do not know.  But I find them interesting.   Here is another.  Dennett~\cite{DennettA} argues, to my mind persuasively, that in discussions of consciousness it is essential to take careful account of the probe (\emph{i.e.}\ the specific question used to elicit a response at a specific time in a specific experimental context).  Furthermore, if one tries to interpret the results obtained using different  probes in terms of a single, coherent story---a ``trajectory of consciousness''---one runs into  difficulties (see, for instance, Dennett's discussion of the colour phi and cutaneous rabbit experiments).  Also, the probe disturbs the system:  it can bring into existence a conscious content which  otherwise might not have occurred.  This is all  reminiscent of the situation in quantum mechanics (there are major differences, but it is reminiscent).

\section{Before Descartes}

At this stage it will be useful to look at the historical development of Cartesian ideas.  In the first place this is a good way to see that the Cartesian concept of consciousness, so far from being a natural intuition (as I believe many people are still inclined to think), actually depends on  postulates which, although they have since become second-nature for many people, originally had to be worked out slowly and laboriously.  In the second place, it brings out the fact that the Cartesian philosophy was intimately related to the  $17^{\mathrm{th}}$ century development of modern science.

The Cartesian concept of consciousness is a $17^{\mathrm{th}}$ century invention.   It did not exist before\footnote{Rorty~\cite{Rorty} makes this point in some detail.  His discussion is very useful.  However, Rorty is not much interested in natural science.  In his own words, he tends to ``view natural science as in the business of controlling and predicting things, and as largely useless for philosophical purposes'' (Saatkamp~\cite{Saatkamp}, p.32).  Consequently he misses a number of points which are crucial for the present discussion. Burtt~\cite{Burtt} is also very  relevant.}. In order to appreciate just how original a departure it was, one needs to see it in the context of the earlier conceptions it replaced.
Concerning  classical Graeco-Roman philosophical ideas\footnote{In the interests of brevity I will here confine myself to the European,  Islamic  and Jewish philosophical traditions, which are  closely related, and which are the ones most relevant to Descartes' intellectual \emph{milieu}.  For the bearing of Buddhism on the problem of consciousness see Blackmore\cite{BlackmoreD}.}  Matson~\cite{Matson} writes
\begin{quote}
Any teaching assistant can set up the mind-body problem so that any freshman will be genuinely worried about it.  Yet none of the ancients ever dreamed of it, not even the author of \emph{De Anima}. 
\end{quote}
and he goes on to observe that  ``In the whole classical corpus there exists no denial of the view that sensing is a bodily process throughout.''   Similarly, Caston~\cite{Caston}, discussing the question whether ``Aristotle even had a concept of consciousness,'' observes that, although ``Aristotle clearly distinguishes being awake and alert from being asleep or knocked out'', he ``does not use any single word to pick out the phenomena we have in mind,''  and he ``does not share the epistemological concerns distinctive of the Cartesian conception of consciousness, such as privacy or indubitability''.  In other words, Aristotle had the everyday concept of consciousness, but not the Cartesian one.  

\subsection{Augustine}
There were philosophers in the ancient Graeco-Roman world whose thinking was in \emph{some} ways similar to the Cartesian philosophy.  The one who came closest was probably St.~Augustine.  It has been suggested, in fact, that Augustine was a significant influence on Descartes~\cite{Rorty,MatthewsA,Menn,CWilson,MatthewsB,Mann}, though opinions differ as to the extent of that influence\footnote{Descartes himself explicitly denied that he had been influenced (though he welcomed what he considered to be the few superficial and purely accidental resemblances as providing useful ammunition in his arguments with  Dutch Calvinists)\cite{CWilson}.  However, as Wilson~\cite{CWilson} points out, that is not, by itself, conclusive since Descartes was in the habit of downplaying, and even outright denying his intellectual debts.}.  Like other philosophers in the Platonic and neo-Platonic tradition (and as one might expect of a Christian theologian) Augustine believed in the existence of an immortal soul.  He also thought that one has indubitable knowledge of one's own existence: 
\begin{quote}
In respect of these truths, I am not at all afraid of the arguments of the Academicians, who say, What if you are deceived?  For if I am deceived, I am.  For he who is not, cannot be deceived; and if I am deceived, by this same token I am.  And since I am if I am deceived, how am I deceived in believing that I am?  for it is certain that I am if I am deceived. [Augustine~\cite{AugustineA}, Book XI, Chapter 26]
\end{quote}
However, this anticipation of Descartes' \emph{cogito ergo sum} should not be allowed to obscure the differences between Augustine and Descartes, which are  considerable.  In the first place Augustine, so far from making the indubitability of one's own existence central to his philosophy, only mentions it   halfway through the \emph{City of God}~\cite{AugustineA} (similarly with the  argument as he gives it in \emph{Against the Academics}~\cite{AugustineB} and \emph{On the Trinity}\cite{AugustineC}).  There is no suggestion that the only thing of which one can be really certain is the existence of one's own consciousness, and that everything else must be deduced from that.  On the contrary, he takes it for granted, as something which does not require demonstration, that in most cases sense-perceptions convey genuine and reliable information about the external world (O'Daly~\cite{ODaly}, p.95).  Concerning this point Matthews~\cite{MatthewsB} says
\begin{quote}
It is, I should say, a singularly important fact about Descartes's \emph{Meditations} that reading them can put one in the grip of what has come to be called ``the problem of the external world.''  \dots There  is no similarly desperate ego-isolation in Augustine. 
\end{quote}

In the second place Augustine's concept of the soul was completely different from the Cartesian one.   For  Augustine the soul is the ``the phenomenon of life in things'' (O'Daly~\cite{ODaly}, p.11).   On this conception  a bird needs a soul in order to fly, quite as much a person needs one in order to think.  Finally, Augustine had a  different theory of sensation from Descartes.     Unlike Descartes, he thought of sensation as an active process, in which ``the soul, as agent of sensation, activates the force of sentience through a fine corporeal medium'' (O'Daly~\cite{ODaly}, p.82).  Thus in vision he thought that rays burst  out of the eye and range abroad, ``so that seeing becomes a kind of visual touching, just as hearing is, so to speak, aural touching'' (\emph{ibid}).   In the Cartesian picture the world is conceived as a sort of spectacle, and the observer as a member of the audience, whose role is purely passive.  In Augustine's conception, by contrast, it is as if the audience climbs onto the stage and walks around among the actors, touching and feeling them.  Given that those are his assumptions I feel that one would not expect him to think in Cartesian terms, of consciousness as an internal movie show.  Unfortunately the obscurities of the texts are such that it is difficult to be sure that he does not.    Matthews~\cite{MatthewsD}  takes the view that 
\begin{quote}
Although commentators have sometimes suggested otherwise, Augustine's theory of sense perception is not representational, if one understands by ``a representational theory of sense perception'' one according to which an image or sense-datum is the direct object of perception.
\end{quote}

Kenny~\cite{KennyA}  thinks that judgment is ``most likely'' correct (p.~215).  Spade~\cite{Spade}, on the other hand, takes a different view.  However, it seems to me that the very fact that there is this scope for disagreement is an indication that Augustine cannot really have been thinking in Cartesian terms.  If someone has \emph{genuinely} caught the Cartesian bug they tend to make it very obvious.  

\subsection{Aquinas}

It was no different in the medieval period. As one would expect medieval philosophers had the everyday concept of consciousness.  Moreover Augustine was one of the most widely read philosophers during the medieval period.  As a consequence, ``it was a commonplace in medieval philosophy that no one can be in doubt about the existence of one's own soul''~\cite[p.253]{Lagerlund}.

Philosophers were also familiar with Avicenna's argument, that it is possible to imagine oneself as a disembodied soul, without sensory experiences.  However, they did not have any of the other notions which go to make up the Cartesian concept of consciousness~\cite{Rorty,KennyA,Lagerlund,MarenbonA,MarenbonB,KennyB}.  The medieval philosopher who is most relevant to the present discussion is Aquinas, since he was the most prominent scholastic philosopher, and consequently the figure most responsible for determining the view which Descartes opposed.  Unlike Augustine, who belonged to the Platonic tradition, Aquinas belonged to the Aristotelian one.

  Nevertheless they had certain things in common.  In the first place Aquinas, like Augustine considered the soul to be ``whatever makes the difference between animate and inanimate objects'' (Kenny~\cite{KennyB}, p.129).  So as Aquinas saw it a tree, or a beetle has a soul, just as a person does.  Moreover the soul is implicated in \emph{every} manifestation of life:  in the act of digesting one's food, or the act of conceiving and bearing a child, no less than in the act of thinking.  In the second place Aquinas, like Augustine and like just about every other medieval philosopher, was primarily interested in those aspects of the soul which make people special.  It is these which go to make up the medieval concept of mind.  
  
  The soul of a beetle is capable of sensation, so sensation was not considered to be something mental.  On the other hand neither a beetle, nor any other non-human living organism can have abstract thoughts or take rational decisions (or so medieval philosophers assumed).  Consequently mind, as medieval philosophers conceived it to be, essentially consists of only two faculties of the soul:  intellect and will (see, for example, Kenny~\cite{KennyB} p.16).  The medieval concept of soul was thus much broader than the Cartesian one, while the medieval concept of mind was much narrower (Descartes, by contrast, identified the concepts of mind and soul).    From the fact that this was the way in which medieval philosophers parcelled up the phenomena, I think it can already be seen that they were rather unlikely to arrive at anything like the Cartesian concept of consciousness.  

For our purposes there are two important differences between Aquinas and Augustine.  The first is that Aquinas, following Aristotle, considered that the soul is the form of the body.  This might be thought a surprising view for someone who, as recently as the last century, could fairly be described as the official philosopher of the Catholic Church~\cite{KennyA}.  How, one might ask, is it to be reconciled with a belief in the immortality of the soul?  The answer is, only with  difficulty (see Kenny~\cite{KennyB} for a critical discussion).  Nevertheless, although Aquinas thought that the soul, like the smile of the Cheshire cat, could survive the death of its body, he also thought that what survives is not the person whose soul it was, and, furthermore, not fully human.  As he put it:
\begin{quote}
\dots but the soul, since it is part of the body of a human being, is not a whole human being, and my soul is not I; so even if a soul gains salvation in another life, that is not I or any human being [translated Kenny~\cite{KennyB}, p.138]
\end{quote}
(it was therefore essential, as Aquinas saw it, that the soul should be re-united with the body on the day of judgment). It might, perhaps, be said that the fact that Aquinas thought that the soul is detachable from the body makes him in some sense a dualist (though I doubt  he would have agreed).  However, his dualism (if ``dualism'' is the right word) is less extreme than that of Descartes (Descartes would not have said that what survives the death of my body is ``not I'').  It could be said that Aquinas' conception of human nature is \emph{earthier} than the Cartesian one.
The second important difference is that Aquinas, unlike Augustine, thought of sensation as a passive process.  However, his conception is no closer than Augustine's to the Cartesian concept of an interior movie show.  As Kenny\cite[p.135]{KennyB}, puts it:
\begin{quote}
In Aquinas' theory there are no intermediaries like sense-data which come between perceiver and perceived. In sensation the sense-faculty does not come into contact with a likeness of the sense-object. Instead, it becomes itself like the sense-object, by taking on the sense-object's form. 
\end{quote}

My aim in giving this brief historical review was to stress the originality of Descartes' conception of consciousness.  If, in over 2000 years of previous philosophical thinking, no one had come up with anything like it, then it follows that, whatever else, the idea cannot be regarded as obvious.  The question now arises:  what led Descartes to make such a radical break with the philosophical past?  It is often suggested that religion, and a consequent belief in the immortality of the soul, is a motive for a dualistic conception of human nature.  That may be so, in many cases.  However, I do not think it can account for Descartes adopting a much more radical version of dualism than his medieval predecessors.  Aquinas, like every other major  medieval  Latin philosopher, was first and foremost a theologian, whereas Descartes' interests where strongly secular, being centred on mathematics, physics and physiology.  If religion was the explanation then, of the two, one would expect it to have been Aquinas who had the more ethereal conception of mind.  Yet in fact it was just the other way around. 

\subsection{Galilei}

Although it is impossible to establish this point conclusively,  there are  reasons for believing that Descartes' real motivation came from Galilean physics.  Galilei  was strongly committed to the Pythagorean idea, that the world is fundamentally mathematical in character\cite{Burtt}.  As he put it in a famous passage from \emph{The Assayer}~\cite{Finocchiaro} (p.183)
\begin{quote}
Philosophy is written in this all-encompassing book that is constantly open before our eyes, that is the universe; but it cannot be understood unless one first learns to understand the language and knows the characters in which it is written. It is written in mathematical language, and its characters are triangles, circles, and other geometrical figures; without these it is humanly impossible to understand a word of it, and one wanders around pointlessly in a dark labyrinth. 
\end{quote}

Of course, the universe does not, at first sight, appear to be a book to be written in the language of mathematics.  Galilei consequently needed to account for all the seemingly non-mathematical, qualitative features of the world, such as colours, sounds and smells, which do not easily fit in with his mathematizing programme.  For that purpose he adopted a doctrine of the ancient atomists~\cite{Furley}, and denied that they are features of objective reality at all, asserting instead that they are somehow produced in the ``sensitive body''~\cite[p.185]{Finocchiaro}: 
\begin{quote}
Accordingly, I say that as soon as I conceive of a corporeal substance or material, I feel indeed drawn by the necessity of also conceiving that it is bounded and has this or that shape; that it is large or small in relation to other things; that it is in this or that location and exists at this or that time; that it moves or stands still; that it touches or does not touch another body; and that it is one, a few, or many. Nor can I, by any stretch of the imagination, separate it from these conditions. However, my mind does not feel forced to regard it as necessarily accompanied by such conditions as the following: that it is white or red, bitter or sweet, noisy or quiet,  and pleasantly or unpleasantly smelling; on the contrary, if we did not have the assistance of our senses, perhaps the intellect and the imagination by themselves would never conceive of them. Thus, from the point of view of the subject in which they seem to inhere, these tastes, odors, colors, etc., are nothing but empty names; rather they inhere only in the sensitive body, such that if one removes the animal, then all these qualities are taken away and annihilated.
\end{quote}
I would argue that this passage marks the actual origin of the Cartesian concept of consciousness.  It is true that Galilei himself did not go into details, as to the nature of the ``sensitive body''.  But I think that once this step had been taken the subsequent development, though not inevitable\footnote{Its lack of inevitability can be seen from, for example, the fact that the ancient atomists\cite{Furley} did not develop a concept of consciousness similar to the Cartesian one.}, became very natural.

It is worth noting that  Galilei  did not attempt to justify the distinction between primary qualities\footnote{The terminology ``primary'' and ``secondary'' is actually due to Locke~\cite{Locke}}, supposed to be objectively real, and secondary qualities, supposed to be in some sense illusory.  Descartes did try to justify it, but his justification is not, to my mind, very convincing.  I believe that Burtt~\cite[p.311]{Burtt} gets it about right when he says that in its first inception\footnote{At a later date one could appeal to the empirical successes of the classical theories apparently based on the doctrine, but not at the time of its first inception.} the doctrine of primary and secondary qualities was ``buttressed
by nothing more than a mathematical apriorism''. 

Subsequently, of course, the primary-secondary distinction played an important role in science since it allowed physicists to dismiss all the ostensibly qualitative features as a problem for philosophers, and to concentrate on the quantitative, mathematical description of nature.  Consciousness, in other words, has been useful to physicists because it has served as a garbage can for all the many things they did not want to have to think about.  However, it is time to ask whether it might have outlived its usefulness.  If the quantum revolution had never happened, one might still be able to make a case for the primary-secondary distinction.  But as it is the quantum revolution did happen, and since then the search for primary qualities consistent with quantum mechanics has been a source of endless difficulties. That being so it is worth asking whether we have any good reason for retaining the notion.

%

In a letter to Mersenne, Descartes~\cite[p.124]{DescartesWorksC}, after saying that Galilei
\begin{quote}
philosophizes much more ably than is usual, in that, so far as he can, he abandons the errors of the Schools and tries to use mathematical methods in the investigation of physical questions
\end{quote}
goes on to complain:
\begin{quote}
But he continually digresses, and he does not take time to explain matters fully.  This, in my view, is a mistake:  it shows that he has not investigated matters in an  orderly way, and has merely sought the explanations for some particular effects, without going into the primary causes in nature.
\end{quote}
It is interesting to observe that Drake~\cite{Drake} says something a little reminiscent of this, concerning Galilei's failure to give an explicit statement of the law of inertia:
\begin{quote}
A modern physicist reading Galilei's writings would share the puzzlement --- I might say the frustration --- experienced by Ernst Mach a century ago, when he searched those works in vain for the general statement that (he felt) ought to be there.  It would become evident to you, as it was to Newton and Mach, that Galilei was in possession of the law of inertia, but you would not then be able to satisfy those historians who demand a clear and complete statement, preferably in print, as a condition of priority.
\end{quote} 
Drake notes that, as a result, the first statement of the law of inertia ``in the form and generality which we accept today'' was given by Descartes\footnote{The history of the law of inertia is complicated.  For a more recent discussion, and a rather different assessment of Galilei's role in its discovery, see, for example, Hooper~\cite{Hooper}.}.
 I  imagine that Descartes would have been equally critical of Galilei's failure to go into details, regarding events inside the ``sensitive body.''  I would suggest that one of his aims  in his early works \emph{The World}\cite{DescartesWorks} and \emph{Treatise on Man}\cite{DescartesWorks}, was to rectify those deficiencies.

\section{The Galilean Core of Descartes' Philosophy}
In the mature form of his philosophy, as represented by \emph{Meditations on First Philosophy}\cite{DescartesWorksB} and \emph{Principles of Philosophy}~\cite{DescartesWorks}, Descartes set out to arrive at demonstratively certain knowledge, ultimately resting on the famous proposition \emph{cogito ergo sum}.   However, an examination of the historical record indicates that this may badly obscure the route by which he was originally led to it.  In his early works \emph{The World} and \emph{Treatise on Man} there is no mention of the \emph{cogito} argument.  Instead these works are entirely devoted to a mechanistic description of the world, conceived along the lines Galilei had previously suggested, and of our relation to it.   Moreover, the treatment is not deductive (as  in his subsequent writings) but avowedly hypothetical:  he is at pains to stress that he is not saying how the world definitely \emph{is}, but only how it conceivably \emph{might be}.  

\emph{The World} and \emph{Treatise on Man} form  part of a larger project, which occupied him during the years 1630 to 1633~\cite{Gaukroger}.  The other parts were either never written, or  have been lost; there is also the possibility that parts were included in subsequent publications.  At all events the works as we have them now, published posthumously, are incomplete.  The reason for this is that at the end of 1633 Descartes learned of Galilei's condemnation by the Inquisition and, not wanting to publish something of which the Church disapproved, he chose ``to suppress it rather than publish it in a mutilated form'' (letter to Mersenne of November 1633, \cite[pp.40f]{DescartesWorksC}. 

In \emph{The World} Descartes begins by making the same distinction between primary and secondary qualities that Galilei does in \emph{The Assayer}.  The fact that he uses one of Galilei's own examples (the tickling sensation produced by a feather) suggests that he was well aware of what Galilei had previously written on the subject.  He then goes  on to give a mechanistic account of the world framed entirely in terms of the Galilean primary qualities of shape, size, position, motion and time.

In \emph{Treatise on Man} Descartes turns to a description of the human body, particularly the brain, conceived as a mechanism.   He ends with a promise to give a description of the ``rational  soul''.  Unfortunately this description is one of the parts of the manuscript which was either never written or  has been lost.  However, since everything he says about the brain is conformable with his later accounts (including the  status of the pineal gland), we may assume that he intended to give an account of the soul which was similarly conformable.  Specifically, we may assume that he intended to say that the soul is a separate, immaterial entity interacting with the brain \emph{via} the pineal gland.  

It is fair to say that what Descartes does in these early works is to flesh out, in much greater detail, Galilei's proposal in \emph{The Assayer}.  However, Descartes also introduces a significant novelty:  in place of Galilei's ``sensitive body'' Descartes locates the secondary qualities in an immaterial soul.  It is impossible to prove, of course, but one may plausibly speculate that it was this---the need to find a home for the secondary qualities---which was the original motivation for the Cartesian soul.

I do not say it was inevitable that Descartes would be led to dualism.   Indeed, his contemporary Thomas Hobbes, in the \emph{Third Set of Objections} (published jointly with the \emph{Meditations}), argued for a completely materialistic conception of human nature\cite{DescartesWorksB}.  However, it does seem to me that, given his opinions about primary and secondary qualities, and given the high value he placed on mathematics, it was very natural for Descartes to take such a view.  It would offend  his Pythagorean sensibilities\footnote{Hobbes was not a mathematician, and is unlikely to have shared Descartes' Pythagorean feelings.  Perhaps that is the reason he could accept the move to full materialism.  Perhaps it is also the reason the ancient Atomists (who were not Pythagoreans either) were not led to the Cartesian concept of consciousness.}  to suppose that, located here and there in the otherwise colourless expanse of mathematical mechanism, there are little brightly painted islands.  It would be equally inconsistent to suppose that, dotted around in the mechanism, there are little islands somehow endowed with  subjective colour experiences.  Since he could not locate colour perceptions inside the physical universe, what else could he do but locate them outside?
%

In his later works, beginning with the \emph{Discourse on the Method}, Descartes~\cite{DescartesWorks} presents his ideas in a very different way.  In particular, the \emph{cogito} argument, which is not mentioned at all in  \emph{The World} and \emph{Treatise on Man}, now becomes central.  This argument is another of Descartes' strikingly original departures from  previous philosophical thinking.  As I mentioned earlier, it was a medieval commonplace, due originally to Augustine, that one cannot doubt the existence of one's own soul~\cite{Lagerlund}.  Moreover there was a widespread interest in sceptical arguments during the early Modern period~\cite{Popkin}.  However, there was no precedent for the way in which Descartes put these  ingredients together.


The \emph{cogito} argument begins with what is sometimes called an act of  hyperbolic doubt.  It is worth asking what motivated this step.  As Wittgenstein\cite{Wittgenstein} has stressed one needs reasons to doubt.  One also needs a suitable context.  At least, one does if one wants people to listen.    Suppose someone expressed doubt, as to whether their head contained sawdust instead of brains (Wittgenstein~\cite[p.36e]{Wittgenstein}).  This would be a much more modest doubt than the global, all-encompassing act of scepticism with which Descartes begins the \emph{cogito} argument.  Yet no one would  take it seriously.  While people \emph{have} taken the Cartesian doubt very seriously indeed:  it is fair to say that the problem of the external world, and  the various philosophical movements to which it has given rise has been the dominant theme in Western philosophy for the last 350 years.  

Why is that?   I think the answer is that, although in the context of everyday life it would be crazy to doubt the existence of external reality, in the context of the views expressed in \emph{The Assayer},  \emph{The World} and \emph{Treatise on Man} the doubt becomes very reasonable.  If one has become convinced that, in sober truth, our senses are radically misleading us as to the existence of colours, sounds, tastes \emph{etc}, then it is surely very natural to wonder if they might also be misleading us as to the existence of shapes, sizes, positions \emph{etc}.  And if one has got as far as wondering if the senses are to be trusted \emph{at all}, then how does one avoid doubting the existence of external reality?  Moreover, I would suggest that that reason for doubting was operative, not only in the mind of Descartes, but also in the minds of his philosophical successors.  It was operative precisely because it was widely believed that science had shown that our senses are radically misleading us.  Scientists who are scornful of philosophical worries about the existence of the external world miss the point:  it was science itself (or what people thought of as science) which originally motivated the worries~\cite{Burtt}.

In short,  I would suggest that all the distinctive features of the Cartesian philosophy are  consequences\footnote{Not consequences in a rigorous, deductive logical sense, but in a looser, psychological sense. This is close to Burtt's~\cite{Burtt} conclusion.} of Galilei's original Pythagorean hypothesis, that the world is fundamentally mathematical in character, and of the related distinction between primary and secondary qualities.  In particular,  this whole way of thinking is rooted in the Galilean-Cartesian concept of matter.  Cartesian consciousness is a secondary concept, parasitic on that.    

\section{Toward a New Philosophy of Nature}
There is an irony in this story.  In the $17^{\mathrm{th}}$ century there was no possibility of finding solid empirical support for the micro-mechanical explanations of such phenomena as colour, or heat, on which the Galilean-Cartesian philosophy was based.  These explanations remained highly speculative until the $19^{\mathrm{th}}$ century when hard evidence  started to accumulate.  Even then progress was  slow, as can be seen from the fact that in the late $19^{\mathrm{th}}$ century controversy about atomism the two sides were equally matched~\cite{AChalmers,Kuhn,Krips,Psillos}.  A nice illustration of this is the fact that in the 1890's Planck, who was subsequently to inaugurate an atomistic view of electromagnetic radiation, was sceptical about atoms, to the extent that Boltzmann could attribute to him the opinion that work on kinetic theory was a ``waste of time and effort'' (Kuhn~\cite[pp.22--3]{Kuhn}; also see Krips~\cite{Krips}). 

 It was only in the  $20^{\mathrm{th}}$ century that the validity of micro-mechanical explanations of the behaviour of matter was established to the satisfaction of every competent physicist.   The irony is that the same advances which finally vindicated micro-mechanical explanations  also cast serious doubt on Galilean-Cartesian assumptions about what  such explanations ought to be like.  Indeed, one of the key papers leading to the general acceptance of atomism (Einstein's 1905 Brownian motion paper\cite{EinsteinA}) was published in the same year, by the same person, as one of the key papers casting doubt on Galilean-Cartesian assumptions (Einstein's 1905 photoelectric paper\cite{EinsteinB}).   

Quantum mechanics challenges the whole Galilean-Cartesian framework.  It is a challenge which has yet to call forth an adequate response.  The Copenhagen Interpretation provides a way of thinking about quantum experiments which is sufficient for the practical needs of working physicists.  But, as its critics point out, it hardly amounts to a coherent philosophy of nature.  Yet, instead of taking the hint from experiment, and trying to move forward, the response of those critics has mostly been to fall back on old, $17^{\mathrm{th}}$ century modes of thought, and to try to find ways of interpreting quantum phenomena which would be consistent with Cartesian assumptions.  

Over half a century ago Pauli described such attempts as ``regressive'' (see, for instance, the letter to Fierz quoted in Gieser~\cite[p.266]{Gieser}).  It seems to me that everything that has happened since tends to confirm that judgment.  What we need to do is to dig up the Galilean-Cartesian foundations and replace them with a different conceptual structure, better adjusted to all we have learned since the year 1900.

The Cartesian philosophy is built on two key principles:  (1) the Pythagorean hypothesis, that there is one true, complete description of the world, expressible in mathematical language and (2) the distinction between primary and secondary qualities.  I believe we ought to abandon both  those principles.  

The idea, naturally suggested by quantum mechanics, that we should dispense with the Pythagorean hypothesis, produces in many people a sense of vertigo.  They fear that letting go of this is tantamount to letting go of the concept of physical reality.   But that merely shows that they are so fixated on the Galilean-Cartesian way of thinking about physical reality that they are unable to envisage  an alternative.

A description is something human.   The ability to give descriptions evolved (presumably) in the palaeolithic, for the purpose of communicating such facts as the location of the nearest source of flint-nodules.  We have a come a long way since then, cognitively speaking.  Nevertheless, the fact is that our modern mathematical descriptions of nature are all expressible in the language of axiomatic set theory, which is a formalization of the naive set theoretic ideas  that palaeolithic hunter-gatherers (presumably) used when sorting their stone tools, negotiating their intricate family relationships, \emph{etc}.  Moreover, our mathematical descriptions comprise  sequences of propositions, just like the verbal communications of  palaeolithic hunter-gatherers.  In short, our mathematical descriptions bear a clear human imprint.   \emph{Conceivably} the universe splits logically, into a  collection of sentence-sized  morsels, each perfectly adapted to human cognitive capacities\footnote{There is \emph{some} overlap here with the discussion in Chapter 1 of  Rorty~\cite{RortyC}.   However, the fact that I agree with Rorty, that the universe is not   a book, should not be taken to imply that I agree with everything else he says in this chapter.}.  But I see no \emph{a priori} reason for assuming that to be the case. 

Our attitude to this question should be empirical.  If Einstein had achieved the same stunning success, with his attempt to explain quantum mechanics in terms of classical field theory,  that he did with general relativity, then there would be reason to take the Pythagorean hypothesis seriously.  But since he did not, and since no one else has either, I think there are grounds for scepticism.  This is not to say that I question the validity of the \emph{partial} descriptions we are able to give.  Nor is to say that I am an anti-realist. It is not even (necessarily) to deny that God is a mathematician.  It is only to say that God is, perhaps, a little more subtle and (dare I say?) interesting than Galilei gave him credit for being.  

Turning to the primary-secondary distinction, it is obvious that colour perceptions are in some sense subjective.  The question is, however, whether they are any more subjective than, for example, the statement that the $\mathbf{E}$ vector at position $\mathbf{r}$ is $3 \mathbf{i} - 4 \mathbf{j} + 7\mathbf{k} \ \mathrm{V} \mathrm{m}^{-1}$---where by ``statement'' I mean the actual ink marks, or the brain states which occur as one reads them.   It is true that a colour-blind person will fail to discriminate two colours which a normally sighted person  sees to be different:  from which it would seem to follow that the colour-blind person has a different visual experience from the normally sighted person. But then it is equally true that a person who measures the electric field intensity to an accuracy of $\pm 1 \ \mathrm{V}\mathrm{m}^{-1}$ will have a different cognitive experience from a person who uses a different instrument to measure it to an accuracy of $\pm 0.1 \ \mathrm{V}\mathrm{m}^{-1}$.

Colour perceptions, being perceptions, are subjective by definition (in a sense).  But then, so are  quantitative thoughts.  Idealists aside, few people are tempted to suppose that, because the belief, that  carbon has proton number 6, is \emph{only} a belief, therefore carbon does not \emph{really} have proton number 6.  No more should one be tempted to suppose that, because the perception of green is \emph{only} a perception, therefore  grass is not \emph{really} green.  

The function of eyes is to acquire information.  Looking at an object is not the same as listening to a verbal description of that object.   But what one acquires by looking is still information, and to that extent it may be regarded as a kind of statement\footnote{Descartes makes an analogy between words and colours at the beginning of \emph{The World}.  However, he fails to draw what I believe to be the correct conclusion}.   Cartesian-minded classical physicists, like Einstein, supposed that the world is completely describable, in terms of fields (or whatever).  Allowing that to be the case, for the sake of argument, it would not follow that the statements of one's visual system are any more subjective than statements made in the approved mathematical language.  What the classical physicist's description says in one way, using the language of fields, the visual system says in another way, using the language of colours.  

To be sure  visual  statements say less---contain less information---than the classical physics description (supposing that it is valid).  But that does not make them subjective.  If one takes some data given to 10 significant figures, and rounds everything off to 3 significant figures, one loses a lot of information.  But the information which remains is no less objective than it was before.
Worrying about the difference between the mathematical description and the description in terms of colours is like worrying about the difference between a description in English and the same description written out in French. Colour qualities are no more in the head---and no less in the head---than the electromagnetic field is in the head.

Discussions of \emph{qualia} are often vitiated by the idea that there are two pictures involved:  one that is coloured (the picture we get from our eyes) and one that is not (the picture we get from physics).  This idea goes back to Descartes, of course, with his talk of colours not ``resembling'' anything in the object.  It is based on a confusion, since neither of these pictures exists.  There is no picture in the head, as we have seen.  Moreover the mathematical descriptions which physics gives us are not pictures either\footnote{It is impossible to imagine the number 3, in the abstract.  Similarly, it is impossible to imagine quantities like vectors.  The electric field vector, for example.}---any more than a verbal description is a picture.  Thinking that colours do not exist in reality because there are no colours in the mathematical description is like thinking that a city is colourless because the verbal description in the guidebook is printed in black and white.

Back in the Palaeolithic, when language first developed, abstract, symbolic descriptions conveyed much less information than the descriptions we get from our eyes.  It was therefore natural to take the visual description to be the standard, or canonical description, against which verbal descriptions were to be judged.  Effectively, reality was identified with the visual description (supplemented with information obtained from the other senses).   However, with  the development of mathematical physics in the $17^{\mathrm{th}}$ century we found an abstract, symbolic mode of description which, unlike ordinary language, was actually superior to the visual description in terms of informational capacity.  It therefore became natural to take the new mathematical description to be the canonical description:  in effect, to identify reality with the mathematical description.   

It seems to me that the lesson of quantum mechanics is that we should  drop the whole idea of there being a canonical description.  Galilei's book metaphor is profoundly misleading.  There is no mathematical description in the sky. The only descriptions around are the ones we humanly construct and which, being human, are necessarily partial.

\section{Conclusion}

To say that there is no canonical description with which reality can be identified is not to deny the existence of reality.  Supposing there to be a canonical description, we have never known it.  Such knowledge of reality as we possess right now is entirely expressed in terms of our ordinary, humanly constructed descriptions. It is not scepticism to suggest that knowledge  so expressed is all we ever will possess.

In this paper I have essentially confined myself to a criticism of  Cartesian philosophy.  To construct an adequate non-Cartesian philosophy would take an enormous amount of work.  However, I believe there is reason to think that if we were to undertake that project it would lead to a conceptual revolution equal in magnitude to the $17^{\mathrm{th}}$ century Cartesian one.     In particular, it might lead to  conceptions of the world, and of human nature, which  differed as much from the Cartesian conceptions as the latter did from medieval conceptions.  So much so that we would, perhaps, no longer want to use the words ``consciousness'' and ``matter'' (except in their everyday senses, of course).   

\section*{Acknowledgements}
The author is  grateful to the Stellenbosch Institute for Advanced Study for their hospitality while carrying out some of the research for this paper.  Research at Perimeter Institute is supported
by the Government of Canada through Industry Canada and by the
Province of Ontario through the Ministry of Research \& Innovation.

\end{document}